\documentclass[aps,twocolumn,showpacs,preprintnumbers,amsmath,amssymb,superscriptaddress,floatfix,nofootinbib]{revtex4}

\usepackage{graphicx}
\usepackage{epsfig}
\usepackage{epstopdf}
\usepackage{hyperref}
\usepackage{amsmath}
\usepackage{amsfonts}
\usepackage{amssymb}

\begin{document}

\title{$f_2(1810)$ as a triangle singularity}
\date{\today}

\author{Ju-Jun Xie}
\affiliation{Institute of Modern Physics, Chinese Academy of
Sciences, Lanzhou 730000, China}

\author{Li-Sheng Geng} \email{lisheng.geng@buaa.edu.cn}
\affiliation{School of Physics and Nuclear Energy Engineering and
International Research Center for Nuclei and Particles in the
Cosmos and Beijing Key Laboratory of Advanced Nuclear Materials and Physics, Beihang University, Beijing 100191, China}

\author{E.~Oset}
\affiliation{Institute of Modern Physics, Chinese Academy of
Sciences, Lanzhou 730000, China}\affiliation{Departamento de
F\'{\i}sica Te\'orica and IFIC, Centro Mixto Universidad de
Valencia-CSIC Institutos de Investigaci\'on de Paterna, Apartado
22085, 46071 Valencia, Spain}

\begin{abstract}

We perform calculations showing that a source producing $K^*
\bar{K}^*$ in $J = 2$ and $L=0$ gives rise to a triangle singularity
at $1810$ MeV with a width of about $200$ MeV from the mechanism
$K^* \to \pi K$ and then $K\bar{K}^*$ merging into the $a_1(1260)$
resonance. We suggest that this is the origin of the present
$f_2(1810)$ resonance and propose to look at the $\pi a_1(1260)$
mode in several reactions to clarify the issue.

\end{abstract}


\maketitle

\section{Introduction}

Triangle singularities, discussed long ago by
Landau~\cite{Landau:1959fi}, are catching the attention of hadron
physicists recently. The large amount of phenomenology gathered in
different facilities studying intermediate energy reactions and
hadron spectra has widened the range of possible cases where triangle singularities are
relevant. In essence the singularities appear from Feynmann diagrams
with three particles in a loop when the particles are placed on
shell and the momenta are collinear. Yet, some condition is also
necessary for the singularity to appear and it is that the
mechanisms reflect a classical problem in which the external
particle $A$ decays into $1$ and $2$, particle $1$ decays into $3$
and an external particle $B$ and then $2$ and $3$ fuse to give an
external particle $C$. This is the content of the Coleman-Norton
theorem~\cite{Coleman:1965xm}. A very simple analytical way to see
when the triangle singularity appears can be seen in
Ref.~\cite{Bayar:2016ftu}, where a critical discussion of the
suggestion made in Refs.~\cite{Guo:2015umn,Liu:2015fea,Guo:2016bkl}
associating the narrow peak of the pentaquark $P_c(4450)$ observed
in the LHCb collaboration~\cite{Aaij:2015tga} to a triangle
singularity is made.

Recent examples of relevant triangle singularities can be seen in
the $\eta(1405) \to \pi a_0(980)$ and $\eta(1405) \to \pi f_0(980)$
decays~\cite{BESIII:2012aa}, the latter one violating isospin, which
is abnormally enhanced due to the triangle
singularities~\cite{Wu:2011yx,Wu:2012pg,Aceti:2012dj}. Another
example is the case of the ``$a_1(1420)$" claimed as a new resonance
by the COMPASS collaboration~\cite{Adolph:2015pws}, which, as
suggested in Ref.~\cite{Liu:2015taa} and shown explicitly in
Refs.~\cite{Ketzer:2015tqa,Aceti:2016yeb}, represents the decay mode
of the $a_1(1260)$ into $\pi f_0(980)$ due to a triangle singularity
coming from the $a_1(1260) \to K^* \bar{K}$, $K^* \to K \pi$ and $K
\bar{K}$ combining to give the $f_0(980)$.

A more recent example is given in Ref.~\cite{Wang:2016dtb}, where
the enhancement in the cross section of the $\gamma p \to K^+
\Lambda(1405)$ reaction around the $\gamma p$ center of mass energy
$W = 2110$ MeV is associated to a triangle singularity stemming from
the formation of a resonance called $N^*(2030)$ that is dynamically
generated from the vector-baryon interaction~\cite{Oset:2009vf}.
This resonance finds support in the $\gamma p \to K^0 \Sigma^+$
reaction close to the $K^* \Lambda$ and $K^* \Sigma$
thresholds~\cite{Ewald:2011gw} (see Ref.~\cite{Ramos:2013wua} for
the theoretical analysis). For the present problem the resonance
decays into $K^* \Sigma$, then $K^* \to \pi K$ and the $\pi \Sigma$
merge to give the $\Lambda(1405)$.

In this work we wish to show that a peak appears precisely at $1810$
MeV due to a process induced by the nearby $f_2(1640)$ going to
$K^*\bar{K}^*$, $K^* \to \pi K$ and $\bar{K}^* K$ merging into the
$a_1(1260)$ resonance. The strength of the peak will have the same
quantum numbers as the resonance from which it comes from, but the
singularity appears at $1810$ MeV, producing a peak that has given
rise to the claim of the $f_2(1810)$
resonance~\cite{Agashe:2014kda}.

The information on the $f_2(1810)$ in the particle data group book
(PDG)~\cite{Agashe:2014kda} is scarce. It has been seen in a few
experiments and the mass and width are quoted as $1815 \pm 12$ MeV
and $197 \pm 22$ MeV, respectively. The decay modes reported are
$\pi \pi$, $K\bar{K}$, $\eta \eta$, $4\pi^0$, and $\gamma \gamma$,
but one finds there $\Gamma_{\eta \eta}/\Gamma_{\rm total} \sim
0.008^{+0.028}_{-0.003}$~\cite{Longacre:1986fh}, $\Gamma_{\pi
\pi}/\Gamma_{4\pi^0} < 0.75$~\cite{Alde:1987ki}, $\Gamma_{4
\pi^0}/\Gamma_{\eta \eta} \sim 0.8 \pm 0.3$~\cite{Alde:1987ki},
$\Gamma_{K^+ K^-}/\Gamma_{\rm total} \sim
0.003^{+0.019}_{-0.002}$~\cite{Longacre:1986fh}, from which one
concludes that the sum of branching fractions of all decay channels,
where the resonance is observed, is only a small fraction of the
total. The main decay mode is still not identified. From the study
presented here we would conclude that the main decay mode of the
peak should be the $\pi a_1(1260)$, which can be seen in the $\pi
\pi \rho$ channel, a decay mode not investigated so far.

We should note that in the present edition of the
PDG~\cite{Olive:2016xmw} the $f_2(1810)$ resonance appears with the
cautionary labels, ``omitted from the summary table" and ``needs
confirmation." The present work will contribute to clarify the
situation.

\section{Formalism and ingredients} \label{sec:formalism}

\begin{figure}[tbp]
\begin{center}
\includegraphics[scale=0.9]{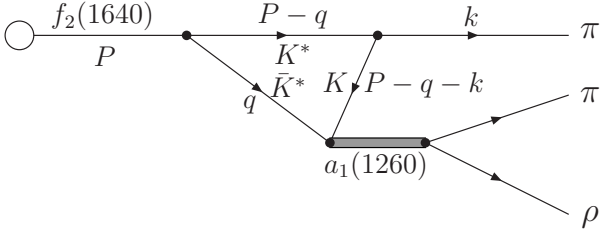}
\caption{Triangle singularity for the production of the $\pi
a_1(1260)$. The circle indicates an external source that produces
the $f_2(1640)$ albeit with an energy different than $1640$ MeV. The
momenta of the particles are also shown.} \label{Fig:fdiagram}
\end{center}
\end{figure}

The tensor mesons have been for a long time an example of a
successful classification in $SU(3)$ multiplets with a $q\bar{q}$
structure~\cite{Giacosa:2005bw,Klempt:2007cp,Crede:2008vw}. However,
the low lying tensor mesons, $f_2(1270)$, $f'_2(1525)$, and
$K^*_2(1430)$ qualify well as dynamically generated states from the
vector-vector interaction~\cite{Molina:2008jw,Geng:2008gx} (see
Ref.~\cite{Xie:2015isa} for a list of reactions supporting this
picture). In this sense, the $f'_2(1525)$ is mostly made from
$K^*\bar{K}^*$~\cite{Geng:2008gx}. If we allow this resonance to
decay into $K^* \bar{K}^*$, then $K^* \to \pi K$ and $\bar{K}^*K \to
a_1(1260)$ (with a mass slightly above the mass threshold of
$\bar{K}^*K$), we find a triangle singularity at $1810$
MeV.~\footnote{It can be easily obtained with the formalism shown in
Eq. (8) of Ref.~\cite{Aceti:2016yeb}. The position of the
singularity in case of zero width for the loop particles can be
easily obtained by means of Eq. (18) of Ref.~\cite{Bayar:2016ftu}.}
Since the width of the $f'_2(1525)$ is only $73$ MeV, the strength of
this resonance at $1810$ MeV is very small and the chances of the
singularity at $1810$ MeV to show up due to the decay of the
$f'_2(1525)$ in that particular channel are very dim. Yet, an
inspection of the PDG shows that in between the $f'_2(1525)$ and
$f_2(1810)$ there are two $f_2$ resonances, the $f_2(1565)$ and
$f_2(1640)$. Due to the proximity of the mass of the $f_2(1640)$ to the
$1810$ MeV, this latter resonance has more chances to influence the
$1810$ MeV region. The $f_2(1640)$ is also not a very well studied
resonance, but the decay modes $\omega \omega$, $4\pi$ and
$K\bar{K}$ have been observed. If the $f_2(1640)$ couples to $\omega
\omega$, then it should also couple to $K^* \bar{K}^*$, which
guarantees that it also decays into $K \bar{K}$ if $K^* \bar{K}^*
\to K \bar{K}$ via $\pi$ exchange. The mechanism that we study is depicted in Fig.~\ref{Fig:fdiagram}.

The coupling of the $f_2(1640)$ has the structure of a
tensor~\cite{Molina:2008jw,Xie:2014twa}, where the polarizations of
$K^*\bar{K}^*$ couple to $J=2$ but in $L=0$. We have
\begin{eqnarray}
 && -iV_{f_2,K^*\bar{K}^*} \equiv  g_{f_2,K^*\bar{K}^*} \times  \nonumber \\
&&\left ( \frac{1}{2}[\epsilon_i(1) \epsilon_j(2) + \epsilon_j(1)
\epsilon_i(2)] - \frac{1}{3} \epsilon_m(1) \epsilon_m(2) \delta_{ij}
\right ), \label{Eq:vf2KstarKbarstar}
\end{eqnarray}
where $\epsilon_i$ are the polarization vectors, $1$ for $K^*$ and
$2$ for $\bar{K}^*$, and only spatial components are considered,
neglecting the three momentum of the vector mesons versus their
masses, as done in Refs.~\cite{Molina:2008jw,Geng:2008gx}. In
Eq.~\eqref{Eq:vf2KstarKbarstar} $g_{f_2,K^*\bar{K}^*}$ is the
coupling constant of the $f_2$ meson to the $K^*\bar{K}^*$ channel.

We also need the coupling of $K^* \to \pi K$ which is easily
obtained from the Lagrangian,
\begin{eqnarray}
{\cal L}_{VPP} = -ig <V^{\mu}[P,\partial_{\mu}P]>,
\end{eqnarray}
where $V$ and $P$ are the $SU(3)$ vector meson matrix and
pseudoscalar meson matrix~\cite{Molina:2008jw,Geng:2008gx},
respectively. The coupling $g$ is,
\begin{eqnarray}
g = \frac{M_V}{2f_{\pi}}, ~~~ M_V = 780 ~{\rm MeV},~~~f_{\pi} = 93
~{\rm MeV}.
\end{eqnarray}

Similarly we need the coupling of the $a_1(1260)$ to $\bar{K}^* K$
and $\pi \rho$, which are a constant times $\vec{\epsilon}(A) \cdot
\vec{\epsilon}(V)$~\cite{Roca:2005nm}, where $\vec{\epsilon}(A)$
stands for the polarization vector of the axial vector meson
$a_1(1260)$ and $\vec{\epsilon}(V)$ for the one of the vector meson
$\bar{K}^*$ or $\rho$. We are only concerned about the shape of the
amplitude compared to a pure $f_2(1640)$ propagator, hence the
couplings do not play a role at this step. Then we have for the
diagram of Fig.~\ref{Fig:fdiagram}
\begin{eqnarray}
t = \frac{1}{P^2 - M^2_{f_2} + i M_{f_2} \Gamma_{f_2}} t_T,
\label{Eq:t}
\end{eqnarray}
where $M_{f_2}$ and $\Gamma_{f_2}$ are the mass and width of the
$f_2(1640)$ meson. In this work we take $M_{f_2} = 1639$ MeV and
$\Gamma_{f_2} = 150$ MeV as in
Refs.~\cite{Agashe:2014kda,Bugg:1995jq}. Meanwhile, $t_T$ is the
triangle amplitude
\begin{eqnarray}
t_T &=& i\int \frac{d^4q}{(2\pi)^4} \frac{1}{q^2 - m^2_{K^*} + i
\epsilon} \frac{1}{(P-q)^2 - m^2_{K^*} + i \epsilon} \times \nonumber \\
&& \frac{1}{(P-q-k)^2 - m^2_{K} + i \epsilon} \left (
\frac{1}{2}[\epsilon_i(1) \epsilon_j(2) + \epsilon_j(1)
\epsilon_i(2)] \right. \nonumber \\
&& \left.  - \frac{1}{3} \epsilon_m(1) \epsilon_m(2) \delta_{ij}
\right )\vec{\epsilon}(1) \cdot (2\vec{k} - \vec{P} + \vec{q})
\vec{\epsilon}(2) \cdot \vec{\epsilon} (A) \times \nonumber \\
&& \frac{1}{(P-k)^2 - M^2_{A} + i M_{A}\Gamma_{A}} \vec{\epsilon}(A)
\cdot \vec{\epsilon} (\rho), \label{Eq:tT}
\end{eqnarray}
where the momenta are shown in Fig.~\ref{Fig:fdiagram} and $M_A$ and
$\Gamma_A$ are the mass and width of the $a_1(1260)$ resonance. We
take $M_A = 1230$ MeV and $\Gamma_A = 425$ MeV as in
Ref.~\cite{Agashe:2014kda}. Besides, $\vec{\epsilon}(\rho)$ stands
for the polarization vector of the $\rho$ meson.

Using the following property
\begin{eqnarray}
\sum_{\rm pol} \vec{\epsilon}_i \vec{\epsilon}_j = \delta_{ij}
\end{eqnarray}
and taking $\vec{P} = 0$ for the $f_2(1640)$ at rest, we find the vertex combination
\begin{eqnarray}
&& \frac{1}{2}  [(2k+q)_i\epsilon_j(\rho) + (2k + q)_j \epsilon_i(\rho) \nonumber \\
&& - \frac{1}{3}(2k + q)_l\epsilon_l(\rho)\delta_{ij}].
\end{eqnarray}

Since when integrating over $\vec{q}$, the only remaining vector is $\vec{k}$, we can write
\begin{eqnarray}
\int d^3\vec{q} ~\! q_i ~\! F(\vec{q}, \vec{k}) \equiv A k_i,
\end{eqnarray}
with $F(\vec{q}, \vec{k})$ the rest of the integrand. From this we
can get
\begin{eqnarray}
A = \int d^3\vec{q} \frac{\vec{q} \cdot \vec{k}}{|\vec{k}|^2}
F(\vec{q}, \vec{k}).
\end{eqnarray}

We also perform analytically the $q^0$ integration as done in Ref.~\cite{Aceti:2015zva} and finally we find
\begin{eqnarray}
t_T = I_2 V_{ij},
\end{eqnarray}
with
\begin{eqnarray}
V_{ij} = k_i \epsilon_j(\rho) + k_j \epsilon_i(\rho) -\frac{2}{3}k_l
\epsilon_l(\rho) \delta_{ij} , \label{Eq:vij}
\end{eqnarray}
and
\begin{eqnarray}
&& I_2 =  \frac{1}{M^2_{\rm inv} - M^2_A + i M_A\Gamma_A} I'_2, \\
&&  I'_2 =  \int \frac{d^3\vec{q}}{(2\pi)^3} (2 + \frac{\vec{q} \cdot \vec{k}}{|\vec{k}|^2}) \times \nonumber \\
&& \frac{1}{8\omega_{\bar{K}^*}(|\vec{q}~|) \omega_{K^*}(|\vec{q}~|) \omega_K(|\vec{q} + \vec{k}|)} \times \nonumber \\
&& \frac{1}{k^0 - \omega_K(|\vec{q} + \vec{k}|) - \omega_{K^*}(|\vec{q}~|)} \times \nonumber \\
&& \frac{1}{P^0 + \omega_{\bar{K}^*}(|\vec{q}~|) + \omega_K(|\vec{q} + \vec{k}|) - k^0} \times \nonumber \\
&& \frac{1}{P^0 - \omega_{\bar{K}^*}(|\vec{q}~|)- \omega_K(|\vec{q} + \vec{k}|) - k^0 + i \Gamma_{K^*}/2} \times \nonumber \\
&& \frac{1}{P^0 - \omega_{K^*}(|\vec{q}~|) - \omega_{\bar{K}^*}(|\vec{q}~|) + i \Gamma_{K^*}} \left\lbrace P^0 \omega_{\bar{K}^*}(|\vec{q}~|) + \right . \nonumber \\
&& \left . k^0\omega_K(|\vec{q} + \vec{k}|) - [\omega_{\bar{K}^*}(|\vec{q}~|) + \omega_K(|\vec{q} + \vec{k}|)] \times \right . \nonumber \\
&& \left . [\omega_{\bar{K}^*}(|\vec{q}~|) + \omega_K(|\vec{q} +
\vec{k}|) + \omega_{K^*}(|\vec{q}~|)] \right\rbrace  ,
\end{eqnarray}
where $\omega_{\bar{K}^*}(|\vec{q}|) = \sqrt{m^2_{K^*} +
|\vec{q}|^2}$, $\omega_{K^*}(|\vec{q}|) = \sqrt{m^2_{K^*} +
|\vec{q}|^2}$, and $\omega_K(|\vec{q} + \vec{k}|) = \sqrt{m^2_{K} +
|\vec{q} + \vec{k}|^2} $ are the energies of $\bar{K}^*$, $K^*$, and
$K$ in the triangle loop, respectively. We take the mass of $K^*$
meson $m_{K^*} =893.1$ MeV and width $\Gamma_{K^*} = 49.1$ MeV. In
addition, $M_{\rm inv}$ is the invariant mass of the $\pi \rho$
system~\footnote{In this work, when we talk about the $\pi \rho$
system, the $\pi$ is always from the decay of the $a_1(1260)$.
Otherwise, the $\pi$ is from the decay of the $K^*$ in the triangle
loop.} decaying from the $a_1(1260)$ resonance, and $|\vec{k}|$ is
the $\pi$ momentum in the $f_2(1640)$ rest frame,
\begin{eqnarray}
|\vec{k}| &=& \frac{\sqrt{[s-(m_{\pi} + M_{\rm inv})^2][s-(m_{\pi} -
M_{\rm inv})^2]}}{2\sqrt{s}},
\end{eqnarray}
with $s = P^2$ the invariant mass squared of the initial $f_2(1640)$
meson.

In order to evaluate $|t|^2$ we must sum over polarizations
\begin{eqnarray}
\sum_{ij} \sum_{\rm pol} V_{ij} V_{ij} = \frac{20}{3} |\vec{k}|^2,
\end{eqnarray}
and then
\begin{eqnarray}
\bar{\sum} \sum |t|^2 = \frac{20 |\vec{k}|^2}{3} |\frac{1}{s -
M^2_{f_2} + i M_{f_2} \Gamma_{f_2}}|^2 |I_2|^2 . \label{Eq:tsquare}
\end{eqnarray}

The differential mass distribution for the $\pi \rho$ decaying from the $a_1(1260)$ is given by
\begin{eqnarray}
\frac{d\Gamma}{dM_{\rm inv}} = \frac{1}{(2\pi)^3}\frac{|\vec{k}|
\tilde{p}_{\rho}}{4s}
 \bar{\sum} \sum |t|^2, \label{Eq:dgdm}
\end{eqnarray}
where $\tilde{p}_{\rho}$ is the $\rho$ momentum in the rest frame of
the $a_1(1260)$ meson,
\begin{eqnarray}
\tilde{p}_{\rho} \!  = \!  \frac{\sqrt{[M^2_{\rm inv} -(m_{\rho} +
m_{\pi})^2][M^2_{\rm inv} -(m_{\rho} - m_{\pi})^2]}}{2M_{\rm inv}},
\end{eqnarray}
with $m_{\rho} = 775.26$ MeV and $m_{\pi} = 138.04$ MeV.

Finally, we obtain $\Gamma$ by integrating Eq.~\eqref{Eq:dgdm} in
$M_{\rm inv}$,
\begin{eqnarray}
\Gamma = \int^{\sqrt{s} - m_{\pi}}_{m_{\rho} + m_{\pi}}
\frac{d\Gamma}{dM_{\rm inv}} dM_{\rm inv}. \label{Eq:Gamma}
\end{eqnarray}
This step allows the contribution of a range of masses for the
$a_1(1260)$ weighted by its spectral function, which is relevant
since the mass of $\bar{K}^* K$ is $1383$ MeV that is above the
nominal mass of the $a_1(1260)$.

\section{Numerical results}

\begin{figure}[htbp]
\begin{center}
\includegraphics[scale=0.45]{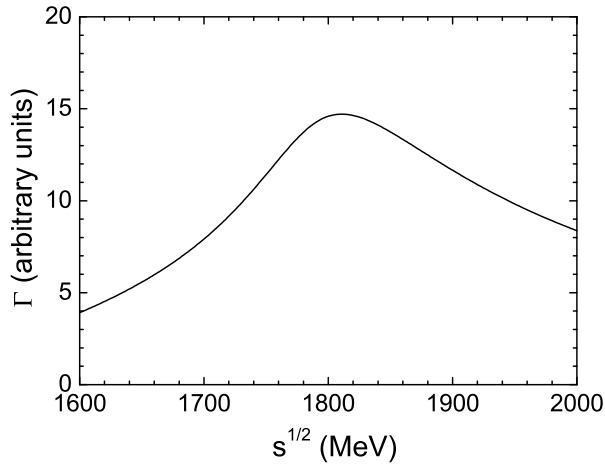}
\caption{$\Gamma$ as a function of $\sqrt{s}$ for Model A without
the $f_2(1640)$ propagator.} \label{Fig:Gamma}
\end{center}
\end{figure}

In Fig.~\ref{Fig:Gamma} we show the results for $\Gamma$ of
Eq.~\eqref{Eq:Gamma} as a function of $\sqrt{s}$, removing the
propagator of the $f_2(1640)$ in Eq.~\eqref{Eq:tsquare} and the
$|\vec{k}|^3$ factor~\footnote{This factor is from the $p$-wave
decay of $f_2(1640) \to \pi a_1(1260)$ and the phase space.} (Model
A). This is done to show the strength of the triangle singularity
alone. We see a broad peak around $1810$ MeV.

Although a triangle singularity gives indeed rise to an infinite
amplitude when the particles in the triangle diagram have zero
width, in practice some of them have width and the amplitude becomes
finite. The explicit consideration of the width of the intermediate
$K^*$, $\bar{K}^*$ and the mass distribution of the $a_1(1260)$
renders the results finite and the singular peak becomes the broad
peak that we observe in Fig.~\ref{Fig:Gamma}.

\begin{figure}[htbp]
\begin{center}
\includegraphics[scale=0.45]{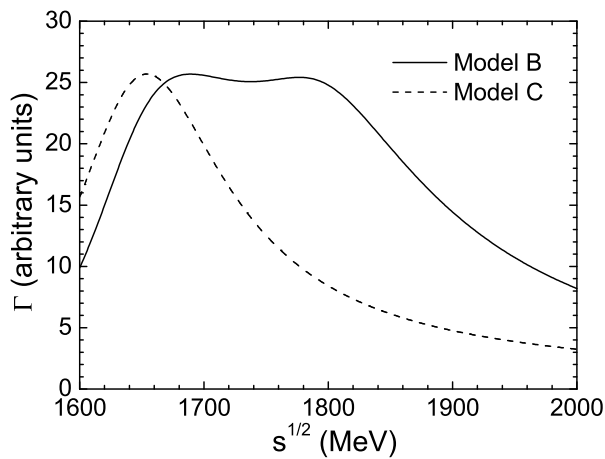}
\caption{$\Gamma$ as a function of $\sqrt{s}$ for Model B (solid
curve) and Model C (dashed curve).} \label{Fig:Gamma-full}
\end{center}
\end{figure}

In Fig.~\ref{Fig:Gamma-full} we plot the full width of
Eq.~\eqref{Eq:Gamma}, taking into account the $f_2(1640)$ propagator
in Eq.~\eqref{Eq:tsquare} and the $|\vec{k}|^3$ factor (Model B).
For comparison we also show the shape of the modulus squared of the
propagator (Model C), removing $t_T$ and normalizing the two curves
to the peak. We can see clear differences between the two curves,
with a large enhancement of the results in the region around $1800$
MeV where the triangle singularity appears. The result with the
solid line gives the shape that we predict if one looks at the decay
mode of the $\pi a_1(1260)$ in the region of $1600-1900$ MeV. A
resonancelike bump shows clearly around $1800$ MeV as a combination
of the $f_2(1640)$ propagator and the structure of the singularity
shown in Fig.~\ref{Fig:Gamma}.

In order to further clarify the issue we take the amplitude
corresponding to the diagram of Fig.~\ref{Fig:fdiagram}, removing
the spin operator $V_{ij}$ of Eq.~\eqref{Eq:vij}. Hence, we consider
the amplitude
\begin{eqnarray}
\tilde{t} = \frac{1}{s - M^2_{f_2} + i M_{f_2} \Gamma_{f_2}} I'_2,
\label{Eq:ttilde}
\end{eqnarray}
in analogy to Eq.~\eqref{Eq:t}, which corresponds to the production of $\pi a_1(1260)$ from an external source with the $a_1(1260)$ treated as
a stable particle.

\begin{figure}[htbp]
\begin{center}
\includegraphics[scale=0.45]{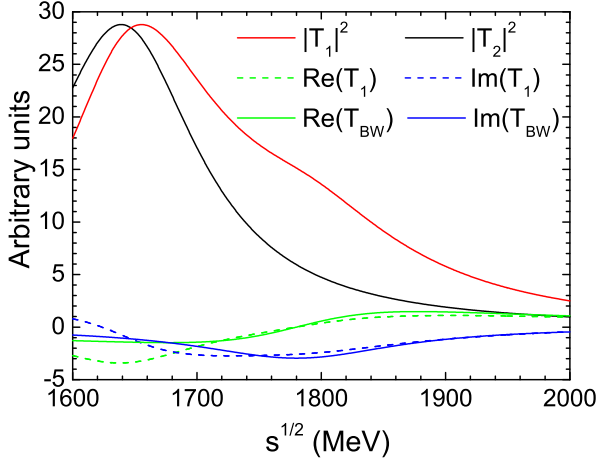}
\caption{Results of $|T_1|^2$, $|T_2|^2$, ${\rm
Re}(T_1)$, ${\rm Im}(T_1)$, ${\rm Re}(T_{\rm BW})$, and ${\rm
Im}(T_{\rm BW})$ as a function of $\sqrt{s}$.} \label{Fig:BW-form}
\end{center}
\end{figure}

We plot the results for $|T_1|^2$ ($T_1 = -i\tilde{t}$) in
Fig.~\ref{Fig:BW-form} with a red-solid curve. In addition, the
modulus squared of the $f_2(1640)$ propagator, $|T_2|^2 =
|\frac{1}{s - M^2_{f_2} + i M_{f_2} \Gamma_{f_2}}|^2$, is shown in
Fig.~\ref{Fig:BW-form} with a black-solid curve, which is normalized
to the peak of $|T_1|^2$. Like in Fig.~\ref{Fig:Gamma-full} we see
the bump corresponding to the ``$f_2(1810)$" resonance. In the
figure we also show the real and imaginary parts of $T_1$ with
green-dashed and blue-dashed curves, respectively. We observe two
structures. Around $\sqrt{s} = 1640$ MeV, looking at $\tilde{t}$
instead of $T_1$, we see the typical Breit-Wigner (BW) structure
that we have introduced by hand in the propagator of $f_2(1640)$ in
Eq.~\eqref{Eq:ttilde}, where ${\rm Im}(\tilde{t})$ [see ${\rm
Re}(T_1)$ in Fig.~\ref{Fig:BW-form}] has a minimum and ${\rm
Re}(\tilde{t})$ [see ${\rm Im}(T_1)$ in Fig.~\ref{Fig:BW-form}]
changes sign at the $f_2(1640)$ resonance mass. Then, when
multiplying $\tilde{t}$ by the phase $-i$ we get a second structure
in $T_1$ that also looks like a BW with a minimum for ${\rm
Im}(T_1)$, and ${\rm Re}(T_1)$ changing sign around $\sqrt{s} =
1780$ MeV. We adopt now an experimental attitude and try to fit this
second structure of $T_1$ by means of a BW amplitude
\begin{eqnarray}
T_{\rm BW} = \frac{\beta}{s- M^2_R + i M_R \Gamma_R}, \label{Eq:tbw}
\end{eqnarray}
where $M_R = 1780$ MeV can be easily obtained at the zero point of
${\rm Re}(T_1)$. We then fit $\beta$ to the value of ${\rm Im}(T_1)$
at $s = M^2_R$ and also get $\Gamma_R$ from the results shown in
Fig.~\ref{Fig:BW-form}, $\Gamma_R \simeq 200$ MeV. Then
\begin{eqnarray}
\beta = - M_R \Gamma_R \times {\rm Im}(T_1)|_{s = M^2_R} .
\end{eqnarray}

The results of ${\rm Re}(T_{\rm BW})$ and ${\rm Im}(T_{\rm BW})$ are
shown in Fig.~\ref{Fig:BW-form} with green-solid and blue-solid
curves. We find an excellent agreement with $T_1$ from $\sqrt{s} =
1750$ MeV on. From the experimental point of view, this would
qualify as a resonance, and the exercise we have done, shows that it
is in the $\pi a_1(1260)$ channel [we have taken the nominal mass of
the $a_1(1260)$ to make Fig.~\ref{Fig:BW-form}] where this
resonant structure shows up. This allows us to trace the behavior to
the triangle singularity. Apart from the fit of the amplitude one
wishes to see the meaning of the structure observed experimentally.
One wishes to associate resonances to states stemming from the
interaction of quarks and
gluons~\cite{Godfrey:1985xj,Entem:2000mq,Giannini:2001kb} or from
the interaction of hadrons making
molecules~\cite{Molina:2008jw,Geng:2008gx,Oller:1997ti,Kaiser:1998fi,Nieves:1999bx,Fariborz:2009wf}.
However, in the present case the structure observed comes from
neither of these. It comes from a singularity in a triangle diagram
that contains $K^* \bar{K}^* K$ in the intermediate states and has
nothing to do with the interaction of quarks or the interaction of
$\pi a_1(1260)$ which is not taken into account in our work.

Since one important aim in hadron physics is to know the origin of
the resonances and their nature, it is important to single out those
cases where a resonant structure can be attributed to a triangle
singularity or a threshold effect. Identifying them is an important
task, and it would be better to exclude them as ``genuine"
resonances to prevent the misleading work of trying to get them from
theories of quarks or hadron interactions.

\section{Conclusions}

We have studied a triangle singularity driven by a source giving
rise to $K^* \bar{K}^*$ in $J=2$, with $K^* \to \pi K$ and $K
\bar{K}^*$ merging to give the axial vector resonance $a_1(1260)$.
We have shown that this triangle singularity gives rise to a
resonancelike structure with a peak at $1810$ MeV and a width of
about $200$ MeV, consistent with the basic properties of the
$f_2(1810)$ listed in the PDG~\cite{Agashe:2014kda}. We have taken
an arbitrary external source and have chosen the nearby $f_2(1640)$
resonance to be the driving element giving rise to $K^*\bar{K}^*$
with the $f_2(1640)$ quantum numbers, the same as the catalogued
$f_2(1810)$ ``resonance." We find a natural explanation in the
singularity to explain the observed peak, and predict that the main
decay mode should be the $\pi a_1(1260)$. This does not contradict
the information on the decay modes of the $f_2(1810)$ tabulated in
the PDG~\cite{Olive:2016xmw} since the few decay modes where it was
observed account for only a small fraction of the total width.

It would be interesting to look at the $\pi a_1(1260)$ channel in
the region of $1600-1900$ MeV in some decay processes to eventually
find the clear structure that we predict in our calculations.
Possible reactions to see this decay mode would be $J/\psi \to \phi
\pi a_1(1260)$,  $J/\psi \to \omega \pi a_1(1260)$, or $J/\psi \to
\gamma \pi a_1(1260)$. Actually, the mode  $J/\psi \to \gamma
f_2(1810)$ with $f_2(1810) \to \eta \eta$ is measured with a
branching ratio $5.4^{+3.5}_{-2.4}\times
10^{-5}$~\cite{Ablikim:2013hq}. According to our discussion, the
mode $J/\psi \to \gamma f_2(1810) \to \gamma \pi a_1(1260)$ should
have a much bigger rate. Actually, the signal for the $f_2(1810)$ in
Ref.~\cite{Ablikim:2013hq} is weak and only extracted through
partial wave analysis with some ambiguities. The detection of the
$\pi a_1(1260)$ mode should show a much clearer signal around $1810$
MeV, according to the results obtained here. We have shown that in
the $\pi a_1(1260)$ mode the amplitude shows indeed a resonant
structure that can be cast into a Breit-Wigner form. Yet, this
structure does not stem from the interaction of quarks or hadrons
but from the triangle diagram that we have discussed, which produce
a kinematical singularity. It is important to know that to
distinguish structures of this type from other ones corresponding to
genuine states that have a dynamical origin from the interaction of
more elementary components. We can only encourage the performance of
experiments like those quoted above that can shed light on the
$f_2(1810)$ and pave the way to investigate other peaks that might
have a similar origin as the one discussed here.

\section*{Acknowledgments}

One of us, E.~O., wishes to acknowledge support from the Chinese
Academy of Sciences in the Program of ``CAS President's International
Fellowship for Visiting Scientists'' (Grant No.\ 2013T2J0012). This
work is partly supported by the National Natural Science Foundation
of China (Grants No.\ 11475227, 11375024, 11522539, 11505158, and
No.\ 11475015.) and the Youth Innovation Promotion Association CAS
(No.\ 2016367). This work is also partly supported by the Spanish
Ministerio de Economia y Competitividad and European FEDER funds
under the Contracts No. FIS2011-28853-C02-01, No. FIS2011-
28853-C02-02, No. FIS2014-57026-REDT, No. FIS2014-51948-C2- 1-P, and
No. FIS2014-51948-C2-2-P, and the Generalitat Valenciana in the program
Prometeo II-2014/068. We acknowledge the support of the European
Community-Research Infrastructure Integrating Activity Study of
Strongly Interacting Matter (acronym HadronPhysics3, Grant Agreement
No. 283286) under the Seventh Framework Programme of the EU.

\bibliographystyle{plain}

\end{document}